\documentclass[aps,onecolumn,12pt,nofootinbib,showpacs]{revtex4}
\usepackage[dvips]{graphicx} \usepackage{amssymb}
\usepackage{psfrag}

\begin{document}
\begin{flushright}
\vspace*{-1.5cm}
BA-04-13\\
\end{flushright}
\vspace{-.1cm}
\title{Reheat temperature in supersymmetric hybrid inflation models}
\author{V. N. {\c S}eno$\breve{\textrm{g}}$uz} \email{nefer@udel.edu}
\author{Q. Shaf\mbox{}i} \email{shafi@bxclu.bartol.udel.edu}
\affiliation{Bartol Research Institute, University of Delaware, Newark, DE
19716, USA} 
\begin{abstract} 
The allowed range of parameters for supersymmetric hybrid inflation and its
extensions are investigated. The lower bound on the reheat temperature $T_r$ in
these models with hierarchical right handed neutrinos is found to be
$3\times10^7$ GeV. ($T_r$ as low as 100 GeV is possible for quasi degenerate
right handed neutrinos.) We also present revised estimates for the scalar spectral index and
the symmetry breaking scale associated with inflation.
\end{abstract}
\pacs{98.80.Cq, 12.60.Jv, 04.65.+e}
\maketitle
%
%
\section{Introduction}
In supersymmetric (SUSY) hybrid inflation models, inflation is
associated with the breaking of a gauge symmetry $G$ to the minimal
supersymmetric standard model (MSSM). The symmetry breaking scale $M$
is fixed by the amplitude of the primordial curvature perturbation,
and turns out to be of order $10^{16}$ GeV, remarkably close, if not
identical, to the supersymmetric grand unification scale \cite{Dvali:1994ms,Lazarides:2001zd}.

A particularly simple and compelling example of $G$ is provided by the standard model gauge
group supplemented by a gauged $U(1)_{B-L}$ symmetry which requires, from anomaly
cancellations, the presence of three right handed neutrinos. Other, more
elaborate examples of $G$ in which SUSY hybrid inflation \cite{Dvali:1994ms} and its extensions, namely
shifted hybrid inflation \cite{Jeannerot:2000sv} and smooth hybrid inflation \cite{Lazarides:1995vr}, 
have been implemented include SO(10) \cite{Kyae:2002hu} and its various subgroups. In the hybrid and
shifted hybrid inflation models, the scalar spectral index $n_s\ge0.98$,
whereas in the smooth case, $n_s\ge0.97$. Furthermore, 
$\textrm{d}n_{s}/\textrm{d}\ln k\le10^{-3}$,
while the tensor to scalar ratio $r$ turns out to be of order $10^{-4}$ or
less \cite{Senoguz:2003zw}.

The main focus of this paper is to estimate the reheat temperature that is
required to generate sufficient lepton asymmetry following
hybrid, shifted or smooth hybrid inflation. 
Although reheating and leptogenesis in these models has previously been addressed
\cite{Lazarides:1997dv,Asaka:1999jb,Pati:2002pe,Senoguz:2003hc}, the soft SUSY
breaking terms and their impact on inflation were not adequately included. It
turns out that the new terms are significant for estimating the minimum reheat
temperature in these models. We also present revised estimates for the 
symmetry breaking scale associated with inflation as well as the scalar spectral index.

The layout of the paper is as follows. In section \ref{chap2} we provide a brief
review of SUSY hybrid inflation and its extensions. We compute the
allowed range of the dimensionless coupling in the superpotential and the
dependence of the spectral index on this coupling, in the presence of
canonical SUGRA corrections and soft SUSY
violating terms.  In section \ref{chap3} we investigate
reheating and the generation of matter following inflation in these models. 
For hierarchical right handed neutrinos, we
obtain a lower bound on the reheat temperature from the observed baryon
asymmetry. We also briefly comment on the resolution of the MSSM $\mu$ problem
and its impact on the reheat temperature.

\section{SUSY Hybrid Inflation Models} \label{chap2}
The SUSY hybrid inflation model \cite{Dvali:1994ms} is realized by the
renormalizable superpotential \cite{Copeland:1994vg}
\begin{equation} \label{super} W_1=\kappa S(\Phi\overline{\Phi}-M^{2})
\end{equation}
\noindent where $\Phi(\overline{\Phi})$ denote a conjugate pair of superfields
transforming as nontrivial representations of some gauge group $G$, $S$ is a
gauge singlet superfield, and $\kappa$ $(>0)$ is a dimensionless coupling.  A
suitable $U(1)$ R-symmetry, under which $W_1$ and $S$ transform the same way,
ensures the uniqueness of this superpotential at the renormalizable level
\cite{Dvali:1994ms}.  In the absence of supersymmetry breaking, the potential
energy minimum corresponds to non-zero (and equal in magnitude) vacuum expectation 
values (vevs) $(=M)$ for the scalar components in $\Phi$ and $\overline{\Phi}$, 
while the vev of $S$ is zero.  (We use the same notation for superfields and their scalar components.)
Thus, $G$ is broken to some subgroup $H$ which, in many interesting models,
coincides with the MSSM gauge group.

In order to realize inflation, the scalar fields $\Phi$,
$\overline{\Phi}$, $S$ must be displayed from their present minima.  For
$|S|>M$, the $\Phi$, $\overline{\Phi}$ vevs both vanish so that the gauge
symmetry is restored, and the tree level potential energy density
$\kappa^{2}M^{4}$ dominates the universe.  With supersymmetry thus broken,
there are radiative corrections from the $\Phi-\overline{\Phi}$ supermultiplets
that provide logarithmic corrections to the potential, with additional contributions
to the inflationary potential arising from $N=1$ supergravity.

With a minimal Kahler potential one contribution
to the inflationary potential is given by \cite{Copeland:1994vg,
Panagiotakopoulos:1997qd,Linde:1997sj} 
\begin{equation} V_{\rm
SUGRA}=\kappa^{2}M^{4}\left[\frac{|S|^{4}}{2\,m^4_P}+\ldots\right]\,,
\end{equation} 
where $m_{P}=2.4\times10^{18}$ GeV is the reduced Planck mass.
There are additional contributions to the potential arising from the soft
SUSY breaking terms. In $N=1$ SUGRA these include the universal scalar masses
equal to $m_{3/2}$ ($\sim$ TeV), the gravitino mass. 
However, their effect on the
inflationary scenario is negligible, as discussed below.
The more important soft term,
ignored as far as we can tell in all earlier calculations,
is $(2-A)m_{3/2}\kappa M^2 S(+\rm{h.c.})$. For convenience, we write this as
$a\,m_{3/2} \kappa M^2 |S|$, where $a\equiv2|2-A|\cos(\arg S+\arg(2-A))$. 
The effective potential is approximately given by
the radiative corrections \cite{Dvali:1994ms} plus the leading SUGRA correction $\kappa^2 M^4 |S|^4/2m^4_P$ and
this soft term:
{\setlength\arraycolsep{2pt}
\begin{eqnarray} \label{potential} V_{1}=\kappa^{2}M^{4}\bigg[1
&+&\frac{\kappa^{2}\mathcal{N}}{32\pi^{2}} \Big(
2\ln\frac{\kappa^{2}|S|^{2}}{\Lambda^{2}}+(z+1)^{2}\ln(1+z^{-1})\nonumber\\
& +&(z-1)^{2}\ln(1-z^{-1})\Big)+\frac{|S|^4}{2m^4_P} \bigg]+ a\,m_{3/2} \kappa M^2|S|\,, \end{eqnarray}}
where $z\equiv|S|^{2}/M^{2}$, $\mathcal{N}$ is the dimensionality of
the $\Phi$, $\overline{\Phi}$ representations, and $\Lambda$ is
a renormalization mass scale.
We perform our numerical calculations using this potential, taking $|a\,m_{3/2}|$=1 TeV,
and using the well known equations in the slow roll approximation: 
The number of e-folds after the comoving scale $l$ has crossed the horizon is
given by 
\begin{equation} \label{efold1}
N_l=\frac{1}{m^2_P}\int^{\sigma_l}_{\sigma_f}\frac{V\rm{d}\sigma}{V'} \,.\end{equation}
Here $\sigma\equiv\sqrt{2}|S|$ is the normalized real scalar
field, $\sigma_l$ is the value of the field at the comoving scale $l$ and $\sigma_f$
is the value of the field at the end of inflation.
The amplitude of the curvature perturbation $\mathcal{R}$ is given by
\begin{equation} \label{perturb}
\mathcal{R}=\frac{1}{2\sqrt{3}\pi m^3_P}\frac{V^{3/2}}{|V'|}\,,
\end{equation}
which we evaluate for the comoving wavenumber $k_0\equiv0.002$ Mpc$^{-1}$.

It is instructive to discuss small and large $\kappa$ limits of Eq. (\ref{potential}). 
For $\kappa\gg10^{-3}$, Eq. (\ref{potential}) becomes
\begin{equation} \label{v1}
V_1\simeq\kappa^{2}M^{4}\left[1+\frac{\kappa^{2}\mathcal{N}}{32\pi^{2}}2\ln\frac{\kappa^{2}|S|^{2}}{\Lambda^{2}}+
\frac{|S|^{4}}{2m^4_P}\right]\,, \end{equation}
to a good approximation. This potential has been analyzed in \cite{Linde:1997sj,Senoguz:2003zw},
and the presence of the SUGRA correction was shown to 
lead to a blue spectrum for $\kappa\gtrsim0.06/\sqrt{\mathcal{N}}$.

For $\kappa\ll10^{-3}$, $|S_0|\simeq M$ where $S_0$ is the value
of the field at $k_0$, i.e. $z\simeq1$.  (Note that due to the flatness of the
potential the last 55 or so e-folds occur with $|S|$ close to $M$.) 
From Eqs. (\ref{potential}, \ref{perturb}), as $z\to1$
\begin{equation} \label{spert}
\mathcal{R}=\frac{2\sqrt{2}\pi}{\sqrt{3}m^3_P}\frac{\kappa^2 M^4}
{\mathcal{N}\ln(2)\kappa^3 M +8\pi^2 \kappa M^5/m^4_P
+4\pi^2 a\,m_{3/2}}\,.
\end{equation}
The denominator of Eq. (\ref{spert}) contains the radiative, SUGRA and the soft
terms respectively.  Comparing them, we see that the radiative term can be
ignored for $\kappa\lesssim10^{-4}$.
 
For a positive soft term ($a>0$), the maximum value of $\mathcal{R}$
as a function of $M$ is found to be 
\begin{equation} \label{rmax}
\mathcal{R}_{\rm{max}}=\frac{1}{2^{7/10}\,3\, 5^{3/2}\, \pi}\left(\frac{\kappa^6 m_P}{a\,m_{3/2}}\right)^{1/5}\,.
\end{equation}
Setting $\mathcal{R}\simeq4.7\times10^{-5}$ \cite{Spergel:2003cb}, 
we find a lower bound on $\kappa$ ($\simeq10^{-5}$).
For larger values of $\kappa$, there are two separate solutions of $M$ for a given $\kappa$. 

For $a<0$, there are again two solutions, but for the solution with
a lower value of $M$, the slope changes sign as the inflaton rolls for $\kappa\lesssim10^{-4}$ and
the inflaton gets trapped in a false vacuum. The second solution in principle
allows $\kappa<10^{-5}$, but this is not very natural since it requires a delicate 
cancellation between two large terms in the denominator of Eq. (\ref{spert}).

There is also a soft mass term $m^2_{3/2} |S|^2$ in the potential, corresponding to
an additional term $8\pi^2 m^2_{3/2} /\kappa M$ in the denominator of Eq. (\ref{spert}).
We have omitted this term, since it is insignificant
for $\kappa\gtrsim10^{-5}$.

\begin{figure}[htb] 
\includegraphics[angle=0, width=12cm]{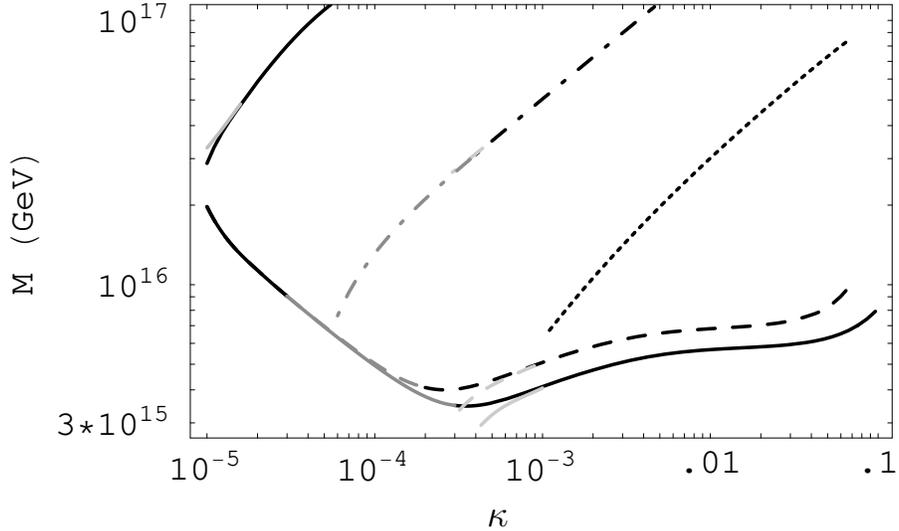} 
\vspace{-1.0cm} 
\begin{center}
{\large \qquad $\kappa$} 
\end{center}
 \vspace{-0.7cm} 
\caption{\sf The value of the symmetry breaking scale $M$ vs. the allowed range of $\kappa$, for SUSY hybrid inflation with $\mathcal{N}=1$ (solid), with $\mathcal{N}=2$ (dashed), and for shifted hybrid inflation 
(dot-dashed for $M_S=m_P$, dotted for $M_S=5\times10^{17}$ GeV). Light grey portions of 
the curves are for $a<0$, where only the segments that do not overlap with the solutions for $a>0$ are shown. 
The grey segments denote the range of $\kappa$ for which the change in $\arg S$ is significant.} \label{fig:21}
\end{figure}

\begin{figure}[hbt] 
\psfrag{k1}{\footnotesize{$\kappa=10^{-4}$}}
\psfrag{k2}{\footnotesize{$\kappa=10^{-3}$}}
\includegraphics[angle=0, width=15cm]{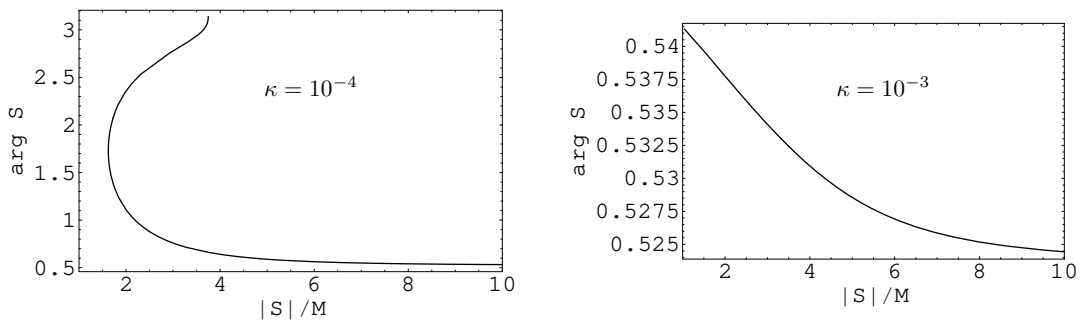} 
 \vspace{-.5cm} 
\caption{\sf Two examples of how $\arg S$ changes as $S$ rolls down, for SUSY hybrid inflation with $\mathcal{N}=2$. 
Left: $\arg S$ exceeds $\pi/2$ before the field reaches the waterfall point,
and the field relaxes in a false vacuum. Right: The field reaches the
waterfall point without a significant change in $\arg S$. 
The initial value of $\arg S=\pi/6$, $\arg(2-A)$ is taken to be zero.} \label{farg}
\end{figure}

The dependence of $M$ on $\kappa$ as well as the allowed range of $\kappa$ is
shown in Fig.  \ref{fig:21}. Note that the soft term depends on $\arg S$, so it
should be checked whether $\arg S$ changes significantly during inflation.
Numerically, we find that it does not, except for a range of $\kappa$ around
$10^{-4}$. For this range, corresponding to the grey segments in the figure,
if the initial value of the $S$ field is greater than $M$ by at least a factor of two or so,
the soft term and the slope become negative even if they were initially positive, before inflation can
suitably end (Fig. \ref{farg}). As mentioned above, 
$|S|\simeq M$ during the last 55 or so e-folds,
so strictly speaking this range is not excluded, although
the required initial conditions may look contrived.

Qualitatively, the duration of inflation is given by $N/H$,
where $N$ denotes the total number of e-folds, and the Hubble parameter
$H\propto\kappa M^2/m_P$. The rate of change in $\arg S$ is $\propto
(m_{3/2}m_P)/|S|$. Therefore one expects that $\arg S$ would change
significantly if $\kappa M^3/N\lesssim m_{3/2}m^2_P$.  For the range of $\kappa$
where the radiative term dominates, $N\propto\kappa^{-2}$ \cite{Linde:1997sj}.
However, for $\kappa\lesssim10^{-4}$, the soft term dominates the slope of the potential,
and Eq. (\ref{spert}) then has solutions with higher values of the slope,
and the duration of inflation is shorter. 
Consequently, $\arg S$ stays fixed also in the left segments of the curves.

\begin{figure}[htb] 
\includegraphics[angle=0, width=12cm]{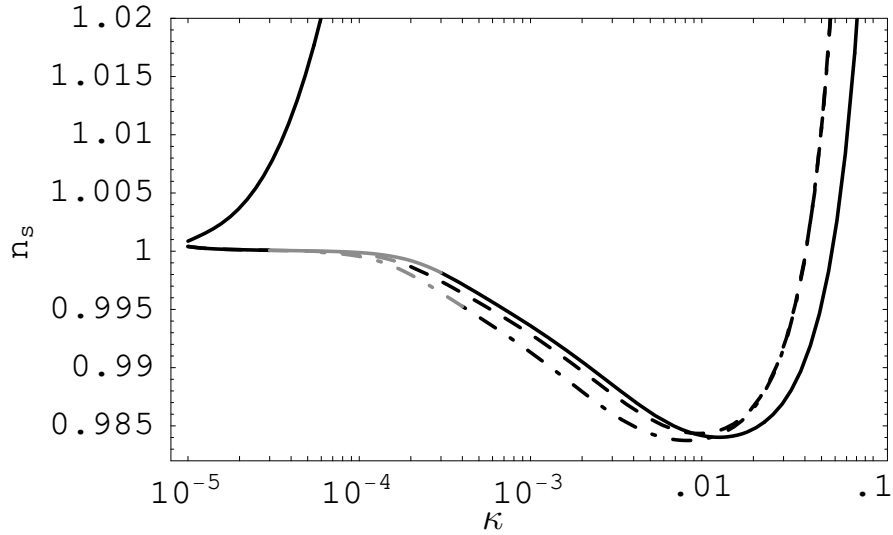} 
\vspace{-1.2cm} 
\begin{center}
{\large \qquad $\kappa$} 
\end{center}
 \vspace{-.5cm} 
\caption{\sf The spectral index $n_s$ vs. the allowed range of $\kappa$, for SUSY hybrid inflation with $\mathcal{N}=1$ (solid), with $\mathcal{N}=2$ (dashed), and for shifted hybrid inflation with $M_S=m_P$ (dot-dashed). The grey segments denote the range of $\kappa$ for which the change in $\arg S$ is significant.} \label{fig:22}
\end{figure}

The dependence of $n_s$ on $\kappa$ is displayed in Fig. \ref{fig:22}.  The segment
with $n_s>1$ for small $\kappa$ corresponds to the solution with a high
symmetry breaking scale.  The running of the spectral index is negligible, with
$\textrm{d}n_{s}/\textrm{d}\ln k\lesssim10^{-3}$. The experimental data
disfavor $n_s$ values in excess of unity on smaller scales (say
$k\lesssim0.05$ Mpc$^{-1}$), which leads us to restrict ourselves to
$\kappa\lesssim0.1/\sqrt{\mathcal{N}}$ for $n_s\le1.04$.\footnote{Larger values of $\kappa$
may be allowed in models where dissipative effects are significant. Such effects
become important for large values of $\kappa$, provided the inflaton also
has strong couplings to matter fields \cite{Bastero-Gil:2004tg}.}
Thus, the vacuum energy density during inflation is considerably smaller than
the symmetry breaking scale. Indeed, the tensor to scalar
ratio $r\lesssim10^{-4}$.

The inflationary scenario based on the superpotential $W_1$ in
Eq. (\ref{super}) has the characteristic feature that the end of
inflation essentially coincides with the gauge symmetry breaking.  Thus,
modifications should be made to $W_1$ if the breaking of $G$ to $H$ leads to
the appearance of topological defects such as monopoles, strings or domain
walls. As shown in \cite{Jeannerot:2000sv}, one simple resolution of the topological defects problem
is achieved by supplementing $W_1$ with a non-renormalizable term:
\begin{equation} \label{super2} W_2=\kappa
S(\overline{\Phi}\Phi-v^{2})-\frac{S(\overline{\Phi}\Phi)^{2}}{M^{2}_{S}}\,,
\end{equation}
\noindent where $v$ is comparable to the SUSY grand unified theory (GUT) scale 
$M_{\rm GUT}\simeq2\times10^{16}$ GeV and $M_{S}$ is an effective
cutoff scale. The dimensionless coefficient of the non-renormalizable term is
absorbed in $M_S$.  The presence of the non-renormalizable term enables an
inflationary trajectory along which the gauge symmetry is broken.  Thus, in
this `shifted' hybrid inflation model the topological defects are inflated away.

The inflationary potential is similar to Eq. (\ref{potential}) \cite{Jeannerot:2000sv}:
{\setlength\arraycolsep{2pt}
\begin{eqnarray} \label{potential2} V_{2}=
\kappa^{2}m^{4}\bigg[1&+&\frac{\kappa^{2}}{16\pi^{2}} \Big(
2\ln\frac{\kappa^{2}|S|^{2}}{\Lambda^{2}}+(z+1)^{2}\ln(1+z^{-1})\nonumber\\
&+&(z-1)^{2}\ln(1-z^{-1})\Big) +\frac{|S|^4}{2m^4_P}\bigg]+a\,m_{3/2}\kappa v^2|S|{}\,.  \end{eqnarray}}
Here $m^{2}=v^{2}(1/4\xi-1)$ with $\xi=v^{2}/\kappa M^{2}_{S}$, $z\equiv2|S|^{2}/m^{2}$,
and $2-A$ is replaced by $2-A+A/2\xi$ in the expression for $a$.
The slow roll parameters (and therefore $n_s$, ${\rm d}n_s/{\rm d}\ln k$, and
$r$) are similar to the SUSY hybrid inflation model (Fig. \ref{fig:22}).

The vev $M$ at the SUSY minimum is given by \cite{Jeannerot:2000sv}
\begin{equation}
\left(\frac{M}{v}\right)^{2}=\frac{1}{2\xi}\left(1-\sqrt{1-4\xi}\right)\,,
\end{equation}
\noindent and is $\sim10^{16}-10^{17}$ GeV depending on $\kappa$ and $M_S$.  
The system follows the inflationary trajectory for
$1/7.2<\xi<1/4$ \cite{Jeannerot:2000sv}, which is satisfied for
$\kappa\gtrsim10^{-5}$ if the effective cutoff scale $M_S=m_P$. 
For lower values of $M_S$, the inflationary trajectory is followed only for 
higher values of $\kappa$, and $M$ is lower for a given $\kappa$ (Fig. \ref{fig:21}). 

A variation on these inflationary scenarios is obtained by imposing a
$Z_{2}$ symmetry on the superpotential, so that only even powers of the
combination $\Phi\overline{\Phi}$ are allowed 
\cite{Lazarides:1995vr,Lazarides:1996rk,Senoguz:2003zw}:
\begin{equation} \label{super3} W_3=S\left(-v^2
+\frac{(\Phi\overline{\Phi})^{2}}{M^{2}_{S}}\right)\,, \end{equation}
\noindent where the dimensionless parameter $\kappa$ is absorbed in $v$.  The resulting scalar
potential possesses two (symmetric) valleys of local minima which are suitable
for inflation and along which the GUT symmetry is broken. As
in the case of shifted hybrid inflation, potential problems associated
with topological defects are avoided.

\begin{figure}[htb] \includegraphics[angle=0,
width=12.5cm]{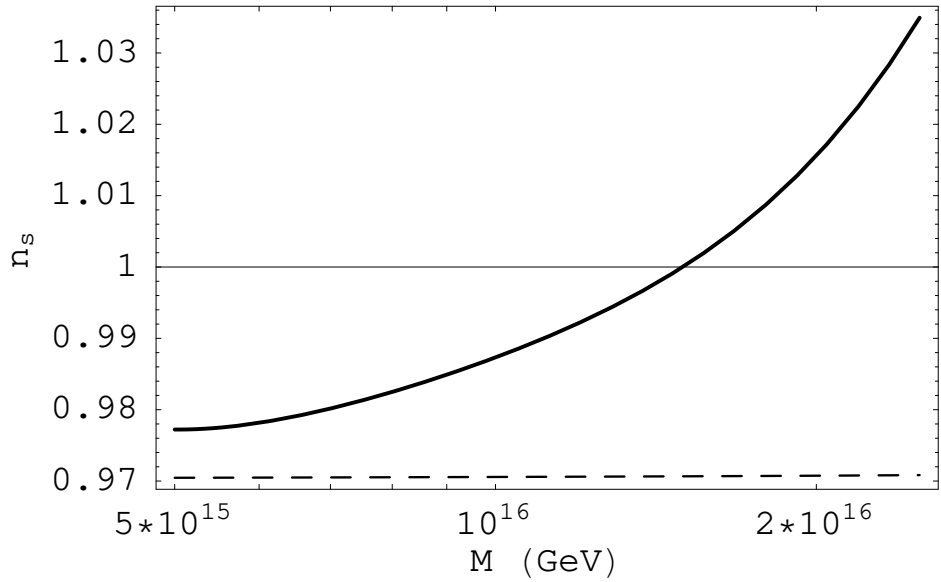} \vspace{-.5cm} \caption{\sf The spectral index $n_s$
as a function of the gauge symmetry breaking
scale $M$ for smooth hybrid inflation (dashed line--without SUGRA
correction, solid line--with SUGRA correction).} \label{fig:24} 
\end{figure}

The vev $M$ at the SUSY minimum is given by
$(v\,M_S)^{1/2}$. For $|S|\gg M$, the inflationary potential
is
\begin{equation} \label{v3}
V_3\approx v^{4}\left[1-\frac{1}{54}\frac{M^4}{|S|^{4}}+\frac{|S|^4}{2m^4_P}\right]\,,
\end{equation}
\noindent where the last term arises from the canonical SUGRA correction.  The
soft terms in this case do not have a significant effect on the inflationary dynamics. In the absence of
the SUGRA correction $n_s\simeq0.97$ \cite{Lazarides:1995vr}. The SUGRA
correction raises $n_{s}$ to above unity for $M\gtrsim1.5\times10^{16}$ GeV,
as shown in Fig. \ref{fig:24} (this figure is slightly different from the figure
published in \cite{Senoguz:2003zw}, due to a computational error in the
latter).

\section{Reheat Temperature and the Gravitino Constraint} \label{chap3}
An important constraint on SUSY hybrid inflation models
arises from considering the reheat temperature $T_{r}$
after inflation, taking into account the gravitino problem which requires that
$T_{r}\lesssim10^6$--$10^{11}$ GeV \cite{Khlopov:1984pf}. This
constraint on $T_r$ depends on the SUSY breaking mechanism and the
gravitino mass $m_{3/2}$. For gravity mediated SUSY breaking models
with unstable gravitinos of mass $m_{3/2}\simeq0.1$--1 TeV,
$T_r\lesssim10^6$--$10^9$ GeV \cite{Kawasaki:1995af}, while
$T_r\lesssim10^{10}$ GeV for stable gravitinos \cite{Fujii:2003nr}.  In gauge
mediated models the reheat temperature is generally more severely constrained, although
$T_r\sim10^9$--$10^{10}$ GeV is possible for $m_{3/2}\simeq5$--100 GeV
\cite{Gherghetta:1998tq}. Finally, the anomaly mediated symmetry breaking (AMSB)
scenario may allow gravitino masses much heavier then a TeV, thus
accommodating a reheat temperature as high as $10^{11}$ GeV \cite{Gherghetta:1999sw}.

After the end of inflation in the models discussed in section~\ref{chap2}, the
fields fall toward the SUSY vacuum and perform damped oscillations about it.
The vevs of $\overline{\Phi}$, $\Phi$ along their right handed neutrino
components $\overline{\nu}^c_H$, $\nu^c_H$  break the gauge symmetry.  The
oscillating system, which we collectively denote as $\chi$, consists of the two
complex scalar fields $(\delta\overline{\nu}^c_H+\delta\nu^c_H)/\sqrt{2}$
(where $\delta\overline{\nu}^c_H$, $\delta\nu^c_H$ are the deviations of
$\overline{\nu}^c_H$, $\nu^c_H$ from $M$) and $S$, with equal mass $m_{\chi}$.

We assume here that the inflaton $\chi$ decays predominantly into right handed neutrino
superfields $N_i$, via the superpotential coupling $(1/m_P)\gamma_{ij}
\overline{\phi}\,\overline{\phi}N_i N_j$ or $\gamma_{ij}\overline{\phi} N_i N_j$,
where $i,j$ are family indices 
(see later for a different scenario connected to the resolution of the MSSM $\mu$
problem). Their subsequent out of equilibrium decay to lepton and Higgs
superfields generates lepton asymmetry, which is then partially converted into
the observed baryon asymmetry by sphaleron effects
\cite{Fukugita:1986hr}. 
The right handed neutrinos, as shown below, can be heavy compared to
the reheat temperature $T_r$.
Without this assumption, the constraints to generate sufficient lepton
asymmetry would be more stringent \cite{Buchmuller:2003gz}.

GUTs typically relate the Dirac neutrino masses to that of the quarks or charged leptons.
It is therefore reasonable to assume the Dirac masses are hierarchical. The
low-energy neutrino data indicates that the right handed neutrinos in this case
will also be hierarchical in general.  As discussed in Ref.
\cite{Akhmedov:2003dg}, setting the Dirac masses strictly equal to the up-type
quark masses and fitting to the neutrino oscillation parameters generally
yields strongly hierarchical right handed neutrino masses ($M_1\ll M_2\ll
M_3$), with $M_1\sim10^5$ GeV. 
The lepton asymmetry in this case is too small
by several orders of magnitude. However, it is plausible that there are large
radiative corrections to the first two family Dirac masses, so that $M_1$ remains 
heavy compared to $T_r$.

\begin{figure}[t] 
\psfrag{m}{\scriptsize{$m_{\chi}$ (GeV)}}
\includegraphics[angle=0, width=12.5cm]{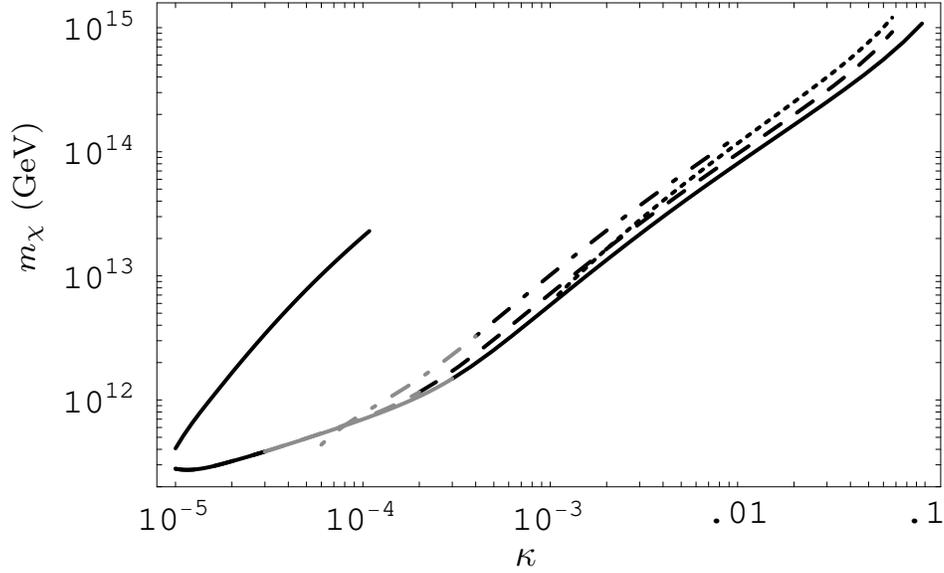} 
\vspace{-0.9cm} 
\begin{center}
{\large \qquad $\kappa$} 
\end{center}
 \vspace{-0.8cm} 
\caption{\sf The inflaton mass $m_{\chi}$ vs. the allowed range of $\kappa$ ($n_s<1.04$),  
for SUSY hybrid inflation with $\mathcal{N}=1$ (solid), with $\mathcal{N}=2$ (dashed), 
and for shifted hybrid inflation 
(dot-dashed for $M_S=m_P$, dotted for $M_S=5\times10^{17}$ GeV). 
The grey segments denote the range of $\kappa$ for which the change in $\arg S$ is significant.} \label{minf}
\end{figure}

\begin{figure}[b] 
\psfrag{m}{\scriptsize{$m_{\chi}$ (GeV)}}
\includegraphics[angle=0, width=12.5cm]{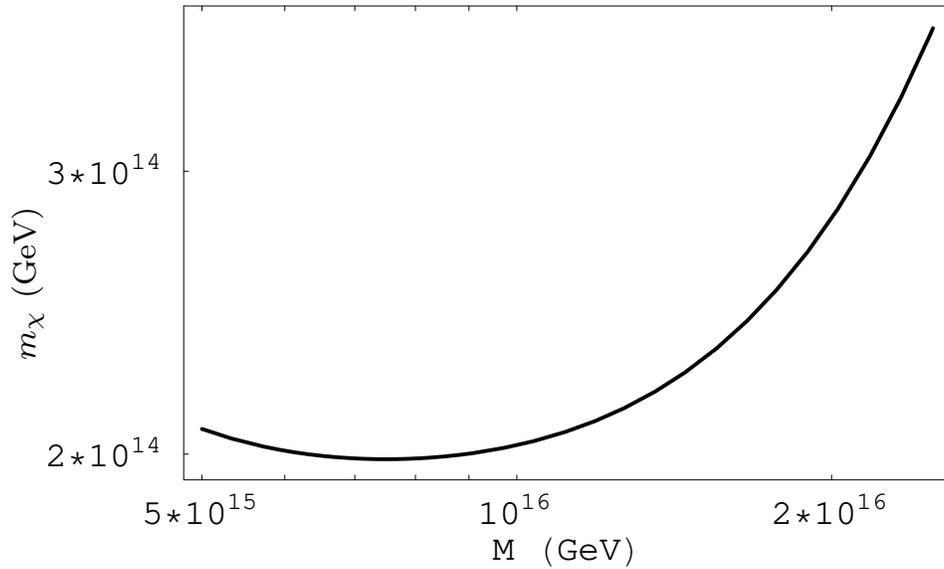} 
 \vspace{-0.8cm} 
\caption{\sf The inflaton mass $m_{\chi}$ vs. the symmetry breaking scale $M$ for smooth hybrid inflation.} \label{minf2}
\end{figure}

\begin{figure}[t] 
\includegraphics[angle=0, width=12cm]{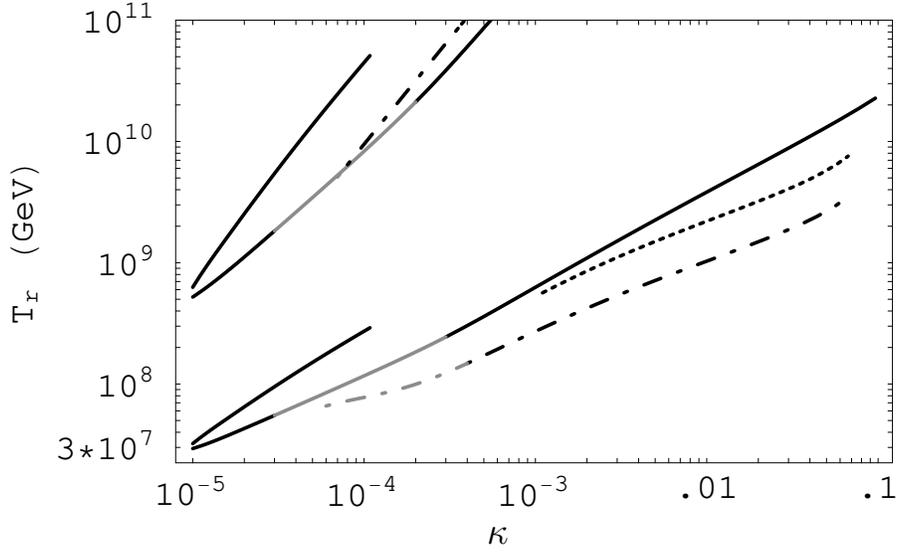} 
\vspace{-1.0cm} 
\begin{center}
{\large \qquad $\kappa$} 
\end{center}
 \vspace{-0.8cm} 
\caption{\sf The lower bound on the reheat temperature $T_r$ vs. the allowed range of $\kappa$ ($n_s<1.04$),  
for SUSY hybrid inflation with $\mathcal{N}=1$ (solid) and for shifted hybrid inflation 
(dot-dashed for $M_S=m_P$, dotted for $M_S=5\times10^{17}$ GeV). The segments in the top left part of the figure correspond to the bounds in the presence of a $\lambda S h^2$ coupling. The grey segments denote the range of $\kappa$ for which the change in $\arg S$ is significant.} \label{ktr}
\end{figure}

\begin{figure}[b] 
\includegraphics[angle=0, width=12cm]{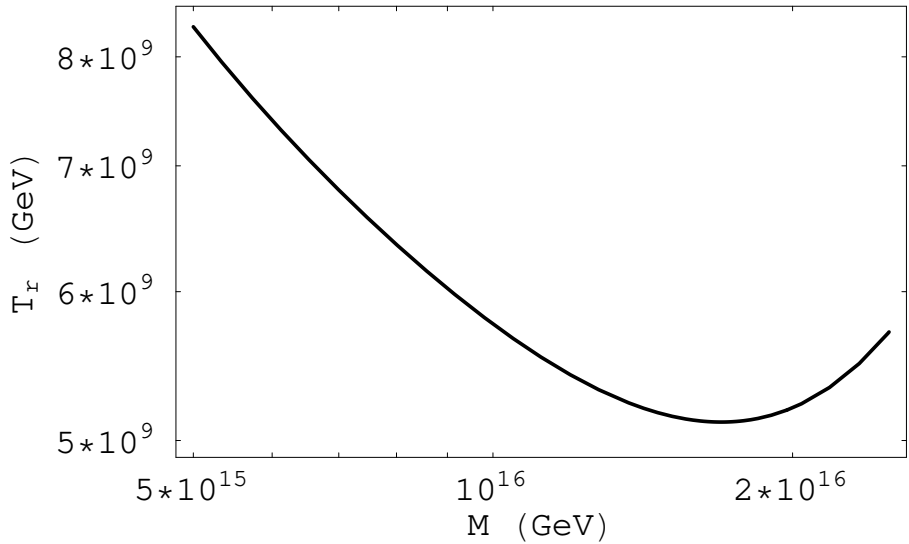}
\vspace{-0.5cm} 
\caption{\sf The lower bound on the reheat temperature $T_r$ vs. the symmetry breaking scale $M$ for smooth hybrid inflation.} \label{sm_mtr}
\end{figure}

A reasonable mass pattern is therefore $M_1<M_2\ll M_3$, which can result from either
the dimensionless couplings $\gamma_{ij}$ or additional symmetries (see e.g. \cite{Senoguz:2004ky}).
The dominant contribution to the lepton asymmetry is
still from the decays with $N_3$ in the loop, as long as the first two family
right handed neutrinos are not quasi degenerate.
Under these assumptions, the lepton asymmetry is given by \cite{Asaka:1999yd}
\begin{equation} \label{nls}
\frac{n_L}{s}\lesssim3\times10^{-10}\frac{T_r}{m_{\chi}}\left(\frac{M_i}{10^6\rm{\
GeV}}\right)\left(\frac{m_{\nu3}}{0.05\rm{\ eV}}\right)\,, \end{equation}
where $M_i$ denotes the mass of the heaviest right handed neutrino the inflaton
can decay into. 
The decay rate $\Gamma_{\chi}=(1/8\pi)(M^2_i/M^2)m_{\chi}$
\cite{Lazarides:1997dv}, and the reheat temperature $T_r$ is given by
\begin{equation} \label{reheat} T_r=\left(\frac{90}{\pi^2
g_*}\right)^{1/4}(\Gamma_\chi\,m_P)^{1/2}\simeq
\frac{1}{10}\frac{(m_P\,m_{\chi})^{1/2}}{M}M_i \,.\end{equation}
(We have ignored the effect
of preheating in hybrid inflation \cite{Garcia-Bellido:1997wm}, which does not seem
to change the perturbative estimate for $T_r$ significantly \cite{Bastero}.)
From the experimental value of the
baryon to photon ratio $\eta_B\simeq6.1\times10^{-10}$ \cite{Spergel:2003cb},
the required lepton asymmetry is found to be $n_L/s\simeq2.5\times10^{-10}$
\cite{Khlebnikov:1988sr}.
Using this value, along with Eqs. (\ref{nls}, \ref{reheat}), we can express $T_r$ in terms of
the symmetry breaking scale $M$ and the inflaton mass $m_{\chi}$:
\begin{equation} \label{trmin}
T_r\gtrsim1.9\times10^{7}{\rm\ GeV}\left(\frac{10^{16}{\rm\
GeV}}{M}\right)^{1/2}\left(\frac{m_{\chi}}{10^{11}{\rm\ GeV}}\right)^{3/4}
\left(\frac{0.05\rm{\ eV}}{m_{\nu3}}\right)^{1/2}\,.  \end{equation}
Here $m_{\chi}$ is given by $\sqrt{2}\kappa M$, $\sqrt{2}\kappa M \sqrt{1-4\xi}$ and
$2\sqrt{2}v^2/M$ respectively for hybrid, shifted hybrid and smooth hybrid inflation.
The value of $m_{\chi}$ is shown in Figs. \ref{minf} and \ref{minf2}.
We show the lower bound on $T_r$ calculated using this equation (taking $m_{\nu3}=0.05$ eV) 
in Figs. \ref{ktr}, \ref{sm_mtr}. 
 
Eq. (\ref{reheat}) also yields the result that the heaviest right handed
neutrino the inflaton can decay into is about 300 (5) times heavier than
$T_r$, for hybrid inflation with $\kappa=10^{-5}$ ($10^{-2}$). For shifted hybrid inflation, this ratio
does not depend on $\kappa$ as strongly and
is $\sim10^2$ \cite{Senoguz:2003hc}. This is consistent with ignoring washout
effects as long as the lightest right handed neutrino mass $M_1$ is also $\gg T_r$.

Both the gravitino constraint and the constraint $M_1\gg T_r$ favor smaller
values of $\kappa$ for hybrid inflation, with
$T_r\gtrsim3\times10^7$ GeV for $\kappa\sim10^{-5}$. 
Similarly, the gravitino constraint favors $\kappa$ values as small as the inflationary trajectory
allows for shifted hybrid inflation, and $T_r\gtrsim10^8$ GeV for $M_S=m_P$.
Smooth hybrid inflation is
relatively disfavored since $T_r\gtrsim5\times10^9$ and $M_2/T_r\simeq3\ (9)$
for $M=5\times10^{15}$ GeV ($2\times10^{16}$ GeV).\footnote{A new inflation model related to
smooth hybrid inflation is discussed in \cite{Senoguz:2004ky}, where
the energy scale of inflation $v$ is lower and consequently lower
reheat temperatures are allowed.}

There are ways to evade these bounds on $T_r$.
Having quasi degenerate neutrinos increases the lepton asymmetry per neutrino
decay $\epsilon$ \cite{Flanz:1996fb} and thus allows lower values of $T_r$ corresponding to lighter right
handed neutrinos. Provided that the neutrino mass splittings are comparable to
their decay widths, $\epsilon$ can be as large as $1/2$ \cite{Pilaftsis:1997jf}. 
The lepton asymmetry in this case is of order $T_r/m_{\chi}$ where $m_{\chi}\sim10^{11}$
GeV for $\kappa\sim10^{-5}$, and sufficient lepton asymmetry 
can be generated with $T_r$ close to the electroweak scale.
For other scenarios that yield $T_r$ of order $10^6$ GeV
in hybrid inflation, see \cite{Asaka:1999jb} (without a $B-L$ symmetry)
and \cite{Antusch:2004hd}.

We end this section with some remarks on the $\mu$ problem
and the relationship to $T_r$ in the present context.
The MSSM $\mu$ problem can naturally be resolved in SUSY hybrid inflation models
in the presence of the term $\lambda Sh^2$ in the superpotential, where
$h$ contains the two Higgs doublets  \cite{Dvali:1998uq}. (The `bare' term $h^2$ is not allowed
by the $U(1)$ R-symmetry.) After inflation
the vev of $S$ generates a $\mu$ term with $\mu=\lambda\langle S\rangle=-m_{3/2}\lambda/\kappa$,
where $\lambda>\kappa$ is required for the required vacuum.
The inflaton in this case predominantly decays into higgses (and higgsinos) with $\Gamma_h=(1/16\pi)
\lambda^2 m_{\chi}$. As a consequence the presence of this term 
significantly increases the reheat temperature $T_r$. Following
\cite{Lazarides:1998qx}, we calculate $T_r$ for the best case scenario
$\lambda=\kappa$. We find  a lower bound on $T_r$ of $5\times10^{8}$ GeV in hybrid inflation,
see Fig. \ref{ktr}. $T_r\gtrsim5\times10^9$ GeV for shifted hybrid inflation with 
$M_S=m_P$\,\footnote{We take $\lambda/\kappa=2/(1/4\xi-1)$ for shifted hybrid inflation. 
Some scalars belonging to the inflaton sector acquire negative mass$^2$
if $\lambda$ is smaller. $\kappa\sim10^{-4}$ corresponds to $\lambda/\kappa\simeq3$.}, 
and $T_r\gtrsim10^{12}$ GeV for smooth hybrid inflation.
An alternative resolution of the $\mu$ problem in these models which has no impact
on $T_r$ invokes an axion symmetry
\cite{Lazarides:1998iq,Jeannerot:2000sv}.

\section{Conclusion}
Supersymmetric hybrid inflation models, through their connection to the grand unification scale, 
provide a compelling framework for the
understanding of the early universe. Such models can also
meet the gravitino and baryogenesis constraints through
non-thermal leptogenesis via inflaton decay.  SUGRA corrections and previously ignored soft SUSY 
violating terms in the inflationary potential lead to lower bounds on $\kappa$ and 
therefore, assuming hierarchical right handed neutrinos, 
on the reheat temperature $T_r$. The lower
bounds on $T_r$ are summarized in Table \ref{ttable}.

\begin{table}[ht]
\caption{Lower bounds on the reheat temperature (GeV)}
\begin{tabular}{|l|r|r|}
\hline
 & without $\lambda S h^2$ & with $\lambda S h^2$ \\
\hline
SUSY hybrid inflation & $3\times10^7$ & $5\times10^8$ \\
\hline
Shifted hybrid inflation & $7\times10^7$ & $5\times10^9$ \\
\hline
Smooth hybrid inflation & $5\times10^9$ & $\gtrsim10^{12}$ \\
\hline
\end{tabular}
\label{ttable}
\end{table}

\begin{acknowledgments}
QS thanks Mar Bastero-Gil for helpful discussions regarding preheating in
hybrid inflation models.
This work is supported by DOE under contract number DE-FG02-91ER40626. 
\end{acknowledgments}

\begin{thebibliography}{25}
\expandafter\ifx\csname natexlab\endcsname\relax\def\natexlab#1{#1}\fi
\expandafter\ifx\csname bibnamefont\endcsname\relax
  \def\bibnamefont#1{#1}\fi
\expandafter\ifx\csname bibfnamefont\endcsname\relax
  \def\bibfnamefont#1{#1}\fi
\expandafter\ifx\csname citenamefont\endcsname\relax
  \def\citenamefont#1{#1}\fi
\expandafter\ifx\csname url\endcsname\relax
  \def\url#1{\texttt{#1}}\fi
\expandafter\ifx\csname urlprefix\endcsname\relax\def\urlprefix{URL }\fi
\providecommand{\bibinfo}[2]{#2}
\providecommand{\eprint}[2][]{\url{#2}}

\bibitem[{\citenamefont{Dvali et~al.}(1994)\citenamefont{Dvali, Shafi, and
  Schaefer}}]{Dvali:1994ms}
\bibinfo{author}{\bibfnamefont{G.~R.} \bibnamefont{Dvali}},
  \bibinfo{author}{\bibfnamefont{Q.}~\bibnamefont{Shafi}}, \bibnamefont{and}
  \bibinfo{author}{\bibfnamefont{R.~K.} \bibnamefont{Schaefer}},
  \bibinfo{journal}{Phys. Rev. Lett.} \textbf{\bibinfo{volume}{73}},
  \bibinfo{pages}{1886} (\bibinfo{year}{1994}), \eprint{hep-ph/9406319}.

\bibitem[{\citenamefont{Lazarides}(2002)}]{Lazarides:2001zd}
For a review and additional references, see
\bibinfo{author}{\bibfnamefont{G.}~\bibnamefont{Lazarides}},
  \bibinfo{journal}{Lect. Notes Phys.} \textbf{\bibinfo{volume}{592}},
  \bibinfo{pages}{351} (\bibinfo{year}{2002}), \eprint{hep-ph/0111328}.

\bibitem[{\citenamefont{Jeannerot et~al.}(2000)\citenamefont{Jeannerot, Khalil,
  Lazarides, and Shafi}}]{Jeannerot:2000sv}
\bibinfo{author}{\bibfnamefont{R.}~\bibnamefont{Jeannerot}},
  \bibinfo{author}{\bibfnamefont{S.}~\bibnamefont{Khalil}},
  \bibinfo{author}{\bibfnamefont{G.}~\bibnamefont{Lazarides}},
  \bibnamefont{and} \bibinfo{author}{\bibfnamefont{Q.}~\bibnamefont{Shafi}},
  \bibinfo{journal}{JHEP} \textbf{\bibinfo{volume}{10}}, \bibinfo{pages}{012}
  (\bibinfo{year}{2000}), \eprint{hep-ph/0002151}.

\bibitem[{\citenamefont{Lazarides and
  Panagiotakopoulos}(1995)}]{Lazarides:1995vr}
\bibinfo{author}{\bibfnamefont{G.}~\bibnamefont{Lazarides}} \bibnamefont{and}
  \bibinfo{author}{\bibfnamefont{C.}~\bibnamefont{Panagiotakopoulos}},
  \bibinfo{journal}{Phys. Rev.} \textbf{\bibinfo{volume}{D52}},
  \bibinfo{pages}{559} (\bibinfo{year}{1995}), \eprint{hep-ph/9506325}.

\bibitem[{\citenamefont{Kyae and Shafi}(2004)}]{Kyae:2002hu}
\bibinfo{author}{\bibfnamefont{B.}~\bibnamefont{Kyae}} \bibnamefont{and}
  \bibinfo{author}{\bibfnamefont{Q.}~\bibnamefont{Shafi}},
  \bibinfo{journal}{Phys. Rev.} \textbf{\bibinfo{volume}{D69}},
  \bibinfo{pages}{046004} (\bibinfo{year}{2004}), \eprint{hep-ph/0212331}.

\bibitem[{\citenamefont{{\c S}eno$\breve{\textrm{g}}$uz and
  Shafi}(2003)}]{Senoguz:2003zw}
\bibinfo{author}{\bibfnamefont{V.~N.} \bibnamefont{{\c
  S}eno$\breve{\textrm{g}}$uz}} \bibnamefont{and}
  \bibinfo{author}{\bibfnamefont{Q.}~\bibnamefont{Shafi}},
  \bibinfo{journal}{Phys. Lett.} \textbf{\bibinfo{volume}{B567}},
  \bibinfo{pages}{79} (\bibinfo{year}{2003}), \eprint{hep-ph/0305089}.

\bibitem[{\citenamefont{Lazarides et~al.}(1997)\citenamefont{Lazarides,
  Schaefer, and Shafi}}]{Lazarides:1997dv}
\bibinfo{author}{\bibfnamefont{G.}~\bibnamefont{Lazarides}},
  \bibinfo{author}{\bibfnamefont{R.~K.} \bibnamefont{Schaefer}},
  \bibnamefont{and} \bibinfo{author}{\bibfnamefont{Q.}~\bibnamefont{Shafi}},
  \bibinfo{journal}{Phys. Rev.} \textbf{\bibinfo{volume}{D56}},
  \bibinfo{pages}{1324} (\bibinfo{year}{1997}), \eprint{hep-ph/9608256}.

\bibitem[{\citenamefont{Asaka et~al.}(2000)\citenamefont{Asaka, Hamaguchi,
  Kawasaki, and Yanagida}}]{Asaka:1999jb}
\bibinfo{author}{\bibfnamefont{T.}~\bibnamefont{Asaka}},
  \bibinfo{author}{\bibfnamefont{K.}~\bibnamefont{Hamaguchi}},
  \bibinfo{author}{\bibfnamefont{M.}~\bibnamefont{Kawasaki}}, \bibnamefont{and}
  \bibinfo{author}{\bibfnamefont{T.}~\bibnamefont{Yanagida}},
  \bibinfo{journal}{Phys. Rev.} \textbf{\bibinfo{volume}{D61}},
  \bibinfo{pages}{083512} (\bibinfo{year}{2000}), \eprint{hep-ph/9907559}.

\bibitem[{\citenamefont{{\c S}eno$\breve{\textrm{g}}$uz and
  Shafi}(2004{\natexlab{a}})}]{Senoguz:2003hc}
\bibinfo{author}{\bibfnamefont{V.~N.} \bibnamefont{{\c
  S}eno$\breve{\textrm{g}}$uz}} \bibnamefont{and}
  \bibinfo{author}{\bibfnamefont{Q.}~\bibnamefont{Shafi}},
  \bibinfo{journal}{Phys. Lett.} \textbf{\bibinfo{volume}{B582}},
  \bibinfo{pages}{6} (\bibinfo{year}{2004}{\natexlab{a}}),
  \eprint{hep-ph/0309134}.

\bibitem[{\citenamefont{Pati}(2002)}]{Pati:2002pe}
\bibinfo{author}{\bibfnamefont{J.~C.} \bibnamefont{Pati}}
  (\bibinfo{year}{2002}), \eprint{hep-ph/0209160}.

\bibitem[{\citenamefont{Copeland et~al.}(1994)\citenamefont{Copeland, Liddle,
  Lyth, Stewart, and Wands}}]{Copeland:1994vg}
\bibinfo{author}{\bibfnamefont{E.~J.} \bibnamefont{Copeland}},
  \bibinfo{author}{\bibfnamefont{A.~R.} \bibnamefont{Liddle}},
  \bibinfo{author}{\bibfnamefont{D.~H.} \bibnamefont{Lyth}},
  \bibinfo{author}{\bibfnamefont{E.~D.} \bibnamefont{Stewart}},
  \bibnamefont{and} \bibinfo{author}{\bibfnamefont{D.}~\bibnamefont{Wands}},
  \bibinfo{journal}{Phys. Rev.} \textbf{\bibinfo{volume}{D49}},
  \bibinfo{pages}{6410} (\bibinfo{year}{1994}), \eprint{astro-ph/9401011}.

\bibitem[{\citenamefont{Panagiotakopoulos}(1997)}]{Panagiotakopoulos:1997qd}
\bibinfo{author}{\bibfnamefont{C.}~\bibnamefont{Panagiotakopoulos}},
  \bibinfo{journal}{Phys. Rev.} \textbf{\bibinfo{volume}{D55}},
  \bibinfo{pages}{7335} (\bibinfo{year}{1997}), \eprint{hep-ph/9702433}.

\bibitem[{\citenamefont{Linde and Riotto}(1997)}]{Linde:1997sj}
\bibinfo{author}{\bibfnamefont{A.~D.} \bibnamefont{Linde}} \bibnamefont{and}
  \bibinfo{author}{\bibfnamefont{A.}~\bibnamefont{Riotto}},
  \bibinfo{journal}{Phys. Rev.} \textbf{\bibinfo{volume}{D56}},
  \bibinfo{pages}{1841} (\bibinfo{year}{1997}), \eprint{hep-ph/9703209}.

\bibitem[{\citenamefont{Spergel et~al.}(2003)}]{Spergel:2003cb}
\bibinfo{author}{\bibfnamefont{D.~N.} \bibnamefont{Spergel}}
  \bibnamefont{et~al.}, \bibinfo{journal}{Astrophys. J. Suppl.}
  \textbf{\bibinfo{volume}{148}}, \bibinfo{pages}{175} (\bibinfo{year}{2003}),
  \eprint{astro-ph/0302209}.

\bibitem[{\citenamefont{Bastero-Gil and Berera}(2004)}]{Bastero-Gil:2004tg}
\bibinfo{author}{\bibfnamefont{M.}~\bibnamefont{Bastero-Gil}} \bibnamefont{and}
  \bibinfo{author}{\bibfnamefont{A.}~\bibnamefont{Berera}}
  (\bibinfo{year}{2004}), \eprint{hep-ph/0411144}.

\bibitem[{\citenamefont{Lazarides et~al.}(1996)\citenamefont{Lazarides,
  Panagiotakopoulos, and Vlachos}}]{Lazarides:1996rk}
\bibinfo{author}{\bibfnamefont{G.}~\bibnamefont{Lazarides}},
  \bibinfo{author}{\bibfnamefont{C.}~\bibnamefont{Panagiotakopoulos}},
  \bibnamefont{and} \bibinfo{author}{\bibfnamefont{N.~D.}
  \bibnamefont{Vlachos}}, \bibinfo{journal}{Phys. Rev.}
  \textbf{\bibinfo{volume}{D54}}, \bibinfo{pages}{1369} (\bibinfo{year}{1996}),
  \eprint{hep-ph/9606297}; 
\bibinfo{author}{\bibfnamefont{R.}~\bibnamefont{Jeannerot}},
  \bibinfo{author}{\bibfnamefont{S.}~\bibnamefont{Khalil}}, \bibnamefont{and}
  \bibinfo{author}{\bibfnamefont{G.}~\bibnamefont{Lazarides}},
  \bibinfo{journal}{Phys. Lett.} \textbf{\bibinfo{volume}{B506}},
  \bibinfo{pages}{344} (\bibinfo{year}{2001}), \eprint{hep-ph/0103229}.

\bibitem[{\citenamefont{Khlopov and Linde}(1984)}]{Khlopov:1984pf}
\bibinfo{author}{\bibfnamefont{M.~Y.} \bibnamefont{Khlopov}} \bibnamefont{and}
  \bibinfo{author}{\bibfnamefont{A.~D.} \bibnamefont{Linde}},
  \bibinfo{journal}{Phys. Lett.} \textbf{\bibinfo{volume}{B138}},
  \bibinfo{pages}{265} (\bibinfo{year}{1984});
\bibinfo{author}{\bibfnamefont{J.~R.} \bibnamefont{Ellis}},
  \bibinfo{author}{\bibfnamefont{J.~E.} \bibnamefont{Kim}}, \bibnamefont{and}
  \bibinfo{author}{\bibfnamefont{D.~V.} \bibnamefont{Nanopoulos}},
  \bibinfo{journal}{Phys. Lett.} \textbf{\bibinfo{volume}{B145}},
  \bibinfo{pages}{181} (\bibinfo{year}{1984}).

\bibitem[{\citenamefont{Kawasaki and Moroi}(1995)}]{Kawasaki:1995af}
\bibinfo{author}{\bibfnamefont{M.}~\bibnamefont{Kawasaki}} \bibnamefont{and}
  \bibinfo{author}{\bibfnamefont{T.}~\bibnamefont{Moroi}},
  \bibinfo{journal}{Prog. Theor. Phys.} \textbf{\bibinfo{volume}{93}},
  \bibinfo{pages}{879} (\bibinfo{year}{1995}), \eprint{hep-ph/9403364};
\bibinfo{author}{\bibfnamefont{R.~H.} \bibnamefont{Cyburt}},
  \bibinfo{author}{\bibfnamefont{J.~R.} \bibnamefont{Ellis}},
  \bibinfo{author}{\bibfnamefont{B.~D.} \bibnamefont{Fields}},
  \bibnamefont{and} \bibinfo{author}{\bibfnamefont{K.~A.} \bibnamefont{Olive}},
  \bibinfo{journal}{Phys. Rev.} \textbf{\bibinfo{volume}{D67}},
  \bibinfo{pages}{103521} (\bibinfo{year}{2003}), \eprint{astro-ph/0211258};
\bibinfo{author}{\bibfnamefont{M.}~\bibnamefont{Kawasaki}},
  \bibinfo{author}{\bibfnamefont{K.}~\bibnamefont{Kohri}}, \bibnamefont{and}
  \bibinfo{author}{\bibfnamefont{T.}~\bibnamefont{Moroi}}
  (\bibinfo{year}{2004}), \eprint{astro-ph/0402490}.

\bibitem[{\citenamefont{Fujii et~al.}(2004)\citenamefont{Fujii, Ibe, and
  Yanagida}}]{Fujii:2003nr}
\bibinfo{author}{\bibfnamefont{M.}~\bibnamefont{Fujii}},
  \bibinfo{author}{\bibfnamefont{M.}~\bibnamefont{Ibe}}, \bibnamefont{and}
  \bibinfo{author}{\bibfnamefont{T.}~\bibnamefont{Yanagida}},
  \bibinfo{journal}{Phys. Lett.} \textbf{\bibinfo{volume}{B579}},
  \bibinfo{pages}{6} (\bibinfo{year}{2004}), \eprint{hep-ph/0310142}.

\bibitem[{\citenamefont{Gherghetta
  et~al.}(1999{\natexlab{a}})\citenamefont{Gherghetta, Giudice, and
  Riotto}}]{Gherghetta:1998tq}
\bibinfo{author}{\bibfnamefont{T.}~\bibnamefont{Gherghetta}},
  \bibinfo{author}{\bibfnamefont{G.~F.} \bibnamefont{Giudice}},
  \bibnamefont{and} \bibinfo{author}{\bibfnamefont{A.}~\bibnamefont{Riotto}},
  \bibinfo{journal}{Phys. Lett.} \textbf{\bibinfo{volume}{B446}},
  \bibinfo{pages}{28} (\bibinfo{year}{1999}{\natexlab{a}}),
  \eprint{hep-ph/9808401}.

\bibitem[{\citenamefont{Gherghetta
  et~al.}(1999{\natexlab{b}})\citenamefont{Gherghetta, Giudice, and
  Wells}}]{Gherghetta:1999sw}
\bibinfo{author}{\bibfnamefont{T.}~\bibnamefont{Gherghetta}},
  \bibinfo{author}{\bibfnamefont{G.~F.} \bibnamefont{Giudice}},
  \bibnamefont{and} \bibinfo{author}{\bibfnamefont{J.~D.} \bibnamefont{Wells}},
  \bibinfo{journal}{Nucl. Phys.} \textbf{\bibinfo{volume}{B559}},
  \bibinfo{pages}{27} (\bibinfo{year}{1999}{\natexlab{b}}),
  \eprint{hep-ph/9904378}.

\bibitem[{\citenamefont{Fukugita and Yanagida}(1986)}]{Fukugita:1986hr}
\bibinfo{author}{\bibfnamefont{M.}~\bibnamefont{Fukugita}} \bibnamefont{and}
  \bibinfo{author}{\bibfnamefont{T.}~\bibnamefont{Yanagida}},
  \bibinfo{journal}{Phys. Lett.} \textbf{\bibinfo{volume}{B174}},
  \bibinfo{pages}{45} (\bibinfo{year}{1986}); 
\bibinfo{author}{\bibfnamefont{G.}~\bibnamefont{Lazarides}} \bibnamefont{and}
  \bibinfo{author}{\bibfnamefont{Q.}~\bibnamefont{Shafi}},
  \bibinfo{journal}{Phys. Lett.} \textbf{\bibinfo{volume}{B258}},
  \bibinfo{pages}{305} (\bibinfo{year}{1991}).

\bibitem[{\citenamefont{Buchmuller et~al.}(2003)\citenamefont{Buchmuller,
  Di~Bari, and Plumacher}}]{Buchmuller:2003gz}
\bibinfo{author}{\bibfnamefont{W.}~\bibnamefont{Buchmuller}},
  \bibinfo{author}{\bibfnamefont{P.}~\bibnamefont{Di~Bari}}, \bibnamefont{and}
  \bibinfo{author}{\bibfnamefont{M.}~\bibnamefont{Plumacher}},
  \bibinfo{journal}{Nucl. Phys.} \textbf{\bibinfo{volume}{B665}},
  \bibinfo{pages}{445} (\bibinfo{year}{2003}), \eprint{hep-ph/0302092}.

\bibitem[{\citenamefont{Akhmedov et~al.}(2003)\citenamefont{Akhmedov, Frigerio,
  and Smirnov}}]{Akhmedov:2003dg}
\bibinfo{author}{\bibfnamefont{E.~K.} \bibnamefont{Akhmedov}},
  \bibinfo{author}{\bibfnamefont{M.}~\bibnamefont{Frigerio}}, \bibnamefont{and}
  \bibinfo{author}{\bibfnamefont{A.~Y.} \bibnamefont{Smirnov}},
  \bibinfo{journal}{JHEP} \textbf{\bibinfo{volume}{09}}, \bibinfo{pages}{021}
  (\bibinfo{year}{2003}), \eprint{hep-ph/0305322}.

\bibitem[{\citenamefont{{\c S}eno$\breve{\textrm{g}}$uz and
  Shafi}(2004{\natexlab{b}})}]{Senoguz:2004ky}
\bibinfo{author}{\bibfnamefont{V.~N.} \bibnamefont{{\c
  S}eno$\breve{\textrm{g}}$uz}} \bibnamefont{and}
  \bibinfo{author}{\bibfnamefont{Q.}~\bibnamefont{Shafi}},
  \bibinfo{journal}{Phys. Lett.} \textbf{\bibinfo{volume}{B596}},
  \bibinfo{pages}{8} (\bibinfo{year}{2004}{\natexlab{b}}),
  \eprint{hep-ph/0403294}.

\bibitem[{\citenamefont{Asaka et~al.}(1999)\citenamefont{Asaka, Hamaguchi,
  Kawasaki, and Yanagida}}]{Asaka:1999yd}
\bibinfo{author}{\bibfnamefont{T.}~\bibnamefont{Asaka}},
  \bibinfo{author}{\bibfnamefont{K.}~\bibnamefont{Hamaguchi}},
  \bibinfo{author}{\bibfnamefont{M.}~\bibnamefont{Kawasaki}}, \bibnamefont{and}
  \bibinfo{author}{\bibfnamefont{T.}~\bibnamefont{Yanagida}},
  \bibinfo{journal}{Phys. Lett.} \textbf{\bibinfo{volume}{B464}},
  \bibinfo{pages}{12} (\bibinfo{year}{1999}), \eprint{hep-ph/9906366}.

\bibitem[{\citenamefont{Garcia-Bellido and
  Linde}(1998)}]{Garcia-Bellido:1997wm}
\bibinfo{author}{\bibfnamefont{J.}~\bibnamefont{Garcia-Bellido}}
  \bibnamefont{and} \bibinfo{author}{\bibfnamefont{A.~D.} \bibnamefont{Linde}},
  \bibinfo{journal}{Phys. Rev.} \textbf{\bibinfo{volume}{D57}},
  \bibinfo{pages}{6075} (\bibinfo{year}{1998}), \eprint{hep-ph/9711360};
\bibinfo{author}{\bibfnamefont{M.}~\bibnamefont{Bastero-Gil}},
  \bibinfo{author}{\bibfnamefont{S.~F.} \bibnamefont{King}}, \bibnamefont{and}
  \bibinfo{author}{\bibfnamefont{J.}~\bibnamefont{Sanderson}},
  \bibinfo{journal}{Phys. Rev.} \textbf{\bibinfo{volume}{D60}},
  \bibinfo{pages}{103517} (\bibinfo{year}{1999}), \eprint{hep-ph/9904315}.

\bibitem{Bastero} M. Bastero-Gil, private communication.

\bibitem[{\citenamefont{Khlebnikov and Shaposhnikov}(1988)}]{Khlebnikov:1988sr}
\bibinfo{author}{\bibfnamefont{S.~Y.} \bibnamefont{Khlebnikov}}
  \bibnamefont{and} \bibinfo{author}{\bibfnamefont{M.~E.}
  \bibnamefont{Shaposhnikov}}, \bibinfo{journal}{Nucl. Phys.}
  \textbf{\bibinfo{volume}{B308}}, \bibinfo{pages}{885} (\bibinfo{year}{1988}).

\bibitem[{\citenamefont{Flanz et~al.}(1996)\citenamefont{Flanz, Paschos,
  Sarkar, and Weiss}}]{Flanz:1996fb}
\bibinfo{author}{\bibfnamefont{M.}~\bibnamefont{Flanz}},
  \bibinfo{author}{\bibfnamefont{E.~A.} \bibnamefont{Paschos}},
  \bibinfo{author}{\bibfnamefont{U.}~\bibnamefont{Sarkar}}, \bibnamefont{and}
  \bibinfo{author}{\bibfnamefont{J.}~\bibnamefont{Weiss}},
  \bibinfo{journal}{Phys. Lett.} \textbf{\bibinfo{volume}{B389}},
  \bibinfo{pages}{693} (\bibinfo{year}{1996}), \eprint{hep-ph/9607310}.

\bibitem[{\citenamefont{Pilaftsis}(1997)}]{Pilaftsis:1997jf}
\bibinfo{author}{\bibfnamefont{A.}~\bibnamefont{Pilaftsis}},
  \bibinfo{journal}{Phys. Rev.} \textbf{\bibinfo{volume}{D56}},
  \bibinfo{pages}{5431} (\bibinfo{year}{1997}), \eprint{hep-ph/9707235};
\bibinfo{author}{\bibfnamefont{A.}~\bibnamefont{Pilaftsis}} \bibnamefont{and}
  \bibinfo{author}{\bibfnamefont{T.~E.~J.} \bibnamefont{Underwood}},
  \bibinfo{journal}{Nucl. Phys.} \textbf{\bibinfo{volume}{B692}},
  \bibinfo{pages}{303} (\bibinfo{year}{2004}), \eprint{hep-ph/0309342}.

\bibitem[{\citenamefont{Antusch et~al.}(2004)\citenamefont{Antusch,
  Bastero-Gil, King, and Shafi}}]{Antusch:2004hd}
\bibinfo{author}{\bibfnamefont{S.}~\bibnamefont{Antusch}},
  \bibinfo{author}{\bibfnamefont{M.}~\bibnamefont{Bastero-Gil}},
  \bibinfo{author}{\bibfnamefont{S.~F.} \bibnamefont{King}}, \bibnamefont{and}
  \bibinfo{author}{\bibfnamefont{Q.}~\bibnamefont{Shafi}}
  (\bibinfo{year}{2004}), \eprint{hep-ph/0411298}.

\bibitem[{\citenamefont{Dvali et~al.}(1998)\citenamefont{Dvali, Lazarides, and
  Shafi}}]{Dvali:1998uq}
\bibinfo{author}{\bibfnamefont{G.~R.} \bibnamefont{Dvali}},
  \bibinfo{author}{\bibfnamefont{G.}~\bibnamefont{Lazarides}},
  \bibnamefont{and} \bibinfo{author}{\bibfnamefont{Q.}~\bibnamefont{Shafi}},
  \bibinfo{journal}{Phys. Lett.} \textbf{\bibinfo{volume}{B424}},
  \bibinfo{pages}{259} (\bibinfo{year}{1998}), \eprint{hep-ph/9710314}; 
\bibinfo{author}{\bibfnamefont{S.~F.} \bibnamefont{King}} \bibnamefont{and}
  \bibinfo{author}{\bibfnamefont{Q.}~\bibnamefont{Shafi}},
  \bibinfo{journal}{Phys. Lett.} \textbf{\bibinfo{volume}{B422}},
  \bibinfo{pages}{135} (\bibinfo{year}{1998}), \eprint{hep-ph/9711288}.

\bibitem[{\citenamefont{Lazarides and Vlachos}(1998)}]{Lazarides:1998qx}
\bibinfo{author}{\bibfnamefont{G.}~\bibnamefont{Lazarides}} \bibnamefont{and}
  \bibinfo{author}{\bibfnamefont{N.~D.} \bibnamefont{Vlachos}},
  \bibinfo{journal}{Phys. Lett.} \textbf{\bibinfo{volume}{B441}},
  \bibinfo{pages}{46} (\bibinfo{year}{1998}), \eprint{hep-ph/9807253}.

\bibitem[{\citenamefont{Lazarides and Shafi}(1998)}]{Lazarides:1998iq}
\bibinfo{author}{\bibfnamefont{G.}~\bibnamefont{Lazarides}} \bibnamefont{and}
  \bibinfo{author}{\bibfnamefont{Q.}~\bibnamefont{Shafi}},
  \bibinfo{journal}{Phys. Rev.} \textbf{\bibinfo{volume}{D58}},
  \bibinfo{pages}{071702} (\bibinfo{year}{1998}), \eprint{hep-ph/9803397}.

\end{thebibliography}
\end{document}